\documentclass[conference]{IEEEtran}
\IEEEoverridecommandlockouts
\usepackage{cite}
\usepackage{algorithm,amsbsy,amsmath,amssymb,epsfig,bbm,mathrsfs,multirow,amsthm,amsfonts}
\usepackage{array}
\usepackage{graphicx}
\usepackage{graphics}
\usepackage{subfigure}
\usepackage{epstopdf}
\usepackage{color}
\usepackage{soul}
\usepackage{bbm}

\usepackage{algpseudocode}
\usepackage{mathtools}
\usepackage{pdfpages}

\clearpage{\pagestyle{empty}\cleardoublepage}
\newcommand{\beq}{\begin{equation}}
\newcommand{\eeq}{\end{equation}}
\newcommand{\bitm}{\begin{itemize}}
\newcommand{\ba}{\begin{array}}
\newcommand{\ea}{\end{array}}
\newcommand{\eitm}{\end{itemize}}
\newcommand{\beqn}{\begin{eqnarray}}
\newcommand{\eeqn}{\end{eqnarray}}
\newcommand{\beqno}{\begin{eqnarray*}}
\newcommand{\eeqno}{\end{eqnarray*}}
\newcommand{\bma}{\begin{displaymath}}
\newcommand{\ema}{\end{displaymath}}
\newcommand{\bnu}{\begin{enumerate}}
\newcommand{\enu}{\end{enumerate}}
\newcommand{\bce}{\begin{center}}
\newcommand{\ece}{\end{center}}
\newcommand{\btb}{\begin{tabular}}
\newcommand{\etb}{\end{tabular}}

\usepackage{graphicx}
\usepackage{textcomp}
\usepackage{xcolor}
\def\BibTeX{{\rm B\kern-.05em{\sc i\kern-.025em b}\kern-.08em
    T\kern-.1667em\lower.7ex\hbox{E}\kern-.125emX}}

\begin{document}
\title{\huge Maximizing the Promptness of Metaverse Systems using Edge Computing by Deep Reinforcement Learning
}

\author{\IEEEauthorblockN{1\textsuperscript{st} Tam Ninh Thi-Thanh}
\IEEEauthorblockA{\textit{Faculty of Information and Communication Technology} \\
\textit{National Academy of Education Management}\\
Hanoi, Vietnam \\
tamntt@niem.edu.vn}
\and
\IEEEauthorblockN{2\textsuperscript{nd} Trinh Van Chien}
\IEEEauthorblockA{\textit{School of Information and Communication Technology} \\
\textit{Hanoi University of Science and Technology}\\
Hanoi, Vietnam \\
chientv@soict.hust.edu.vn}
\and
\IEEEauthorblockN{3\textsuperscript{rd} Hung Tran}
\IEEEauthorblockA{\textit{Faculty of Information Technology} \\
\textit{Phenikaa University}\\
Hanoi, Vietnam \\
hung.tran@phenikaa-uni.edu.vn}
\and
\IEEEauthorblockN{4\textsuperscript{th} Nguyen Hoai Son}
\IEEEauthorblockA{\textit{Faculty of Information Technology} \\
\textit{University of Engineering and Technology, VNU}\\
Hanoi, Vietnam \\
sonnh@vnu.edu.vn}
\and
\IEEEauthorblockN{5\textsuperscript{th} Van Nhan Vo}
\IEEEauthorblockA{\textit{Faculty of Information Technology} \\
\textit{Duy Tan University}\\
Da Nang, Vietnam \\
vonhanvan@dtu.edu.vn}
}

\maketitle
\begin{abstract} Metaverse and Digital Twin (DT) have attracted much academic and industrial attraction to approach the future digital world. This paper introduces the advantages of deep reinforcement learning (DRL) in assisting Metaverse system-based Digital Twin. In this system, we assume that it includes several Metaverse User devices collecting data from the real world to transfer it into the virtual world, a Metaverse Virtual Access Point (MVAP) undertaking the processing of data, and an edge computing server that receives the offloading data from the MVAP. The proposed model works under a dynamic environment with various parameters changing over time. The experiment results show that our proposed DRL algorithm is suitable for offloading tasks to ensure the promptness of DT in a dynamic environment.
\end{abstract}

\thispagestyle{empty}
\section{INTRODUCTION}
\label{sec:introduction}
In machine learning, reinforcement learning (RL) is an approach where an agent learns to make optimal decisions by exploring and interacting with a specific environment. Using a trial-and-error method, the agent determines subsequent actions by considering the rewards of current actions \cite{sutton2018reinforcement}. The aim of RL is for the agent to learn a policy, which is a mapping from states of the environment to actions that maximize cumulative rewards. The five standard elements in RL are the environment, state, reward, policy, and value. The environment represents the physical or simulated world where the agent operates. The state denotes the agent's current condition. Rewards are positive or negative values that provide feedback from the environment to the agent. The policy is a strategy that maps states to actions, guiding the agent to future rewards based on its current state. RL is developed on the mathematical framework of the Markov Decision Process (MDP), which comprises states $\mathcal{S}$, actions $\mathcal{A}$, transition probabilities $\mathcal{P}r$, rewards $\mathcal{R}$, and a discount factor $\gamma$.

On the other hand, the most commonly utilized RL algorithms are Q-learning (QL) and State-Action-Reward-State-Action (SARSA). Numerous authors have applied RL to various applications in wireless communication systems. For instance, \cite{ali2021reinforcement} designed and analyzed the capabilities of RL-enhanced modern communication systems, focusing particularly on efficient spectrum access based on IEEE 802.11 WLAN. In vehicle control, A. Feher et al. \cite{feher2018q} employed Q-learning for lane keeping, while Y. Shan et al. \cite{shan2020reinforcement} used RL for path tracking in autonomous vehicles. However, due to the limitation of the Q-table, which cannot accommodate large action spaces, Deep Reinforcement Learning (DRL) algorithms have been proposed to address this issue. Building on QL and SARSA, advanced neural network-based algorithms such as Deep Q Network (DQN), Double Deep Q Network (DDQN), Deep Deterministic Policy Gradient (DDPG), Soft Actor-Critic (SAC), and Proximal Policy Optimization (PPO) have been developed to optimize agent decision-making and enhance capabilities. The key difference between RL and DRL is that while RL focuses on feedback to achieve the highest reward in low-complexity environments and simple tasks, DRL extends this by incorporating deep neural networks to handle high-complexity environments and more complicated tasks.

The Metaverse is a virtual world where humans, represented as avatars, interact with each other in a three-dimensional space that mimics reality. It is built on the convergence of technologies that enable multisensory interactions with virtual environments, digital objects, and other people, including virtual reality (VR) and augmented reality (AR). The Metaverse comprises four key components: Virtual Worlds, which are immersive 3D environments where users can interact with each other and their surroundings in real-time; Augmented Reality (AR), which enhances the real world through digital overlays; Virtual Reality (VR), offering fully immersive experiences that exclude the physical world; and Mixed Reality (MR), which combines elements of both AR and VR to allow real and virtual elements to interact seamlessly.

Introduced over a decade ago, Digital Twins (DT) have garnered significant attention from both academia \cite{wu2021digital} and industry \cite{tao2018digital}. A digital twin is a virtual representation of a physical object, system, or process, designed to simulate and predict performance characteristics in real-time. It serves as a bridge between the physical and digital worlds, representing the natural world for numerous applications and acting as a virtual implementation of actual physical systems. Achieving the desired functionality of a DT is challenging, as it requires highly accurate real-time synchronization between the features of the real-world system and its digital counterpart. Additionally, many essential properties of DTs rely on networking, including similarity, promptness, scalability, trustworthiness, composability, and flexibility\cite{9120192}.

The integration of the Metaverse and Digital Twins holds significant potential, enabling immersive simulations of real-world systems, enhancing user interaction with Digital Twins in a more intuitive and immersive manner, and improving decision-making through real-time data and simulations that provide valuable insights. Digital twins facilitate seamless interactions between digital and physical objects, acting as a bridge to the Metaverse. As we look toward the future of telecommunications and digital technologies, concepts like Digital Twins and the Metaverse are gaining increasing attention, particularly in the context of 6G networks.

This paper examines task offloading in a DT to serve Metaverse user devices. Specifically, assuming a 6G system context, where latency is extremely low and throughput is extremely high, we focus on the property called "promptness" as presented in previous studies \cite{9863238} and \cite{10122221}. Promptness refers to the quality of being quick or timely in action or response. In the context of Digital Twins serving Metaverse user devices, promptness implies the ability to rapidly and efficiently initiate and conclude actions or processes in accordance with user requests or system requirements. It entails minimizing delays and ensuring swift responsiveness to ensure seamless user experiences and effective utilization of resources.
The main contributions of our paper are outlined as follows:
\begin{itemize}
    \item We successfully applied the proposed deep reinforcement learning algorithm to ensure the promptness of Digital Twins in serving Metaverse User Devices.
    \item We have also conducted simulations to evaluate the performance of our proposed algorithms.
\end{itemize}


\section{Model architecture}

\begin{figure*}
\centering
\includegraphics[width=0.7\linewidth]{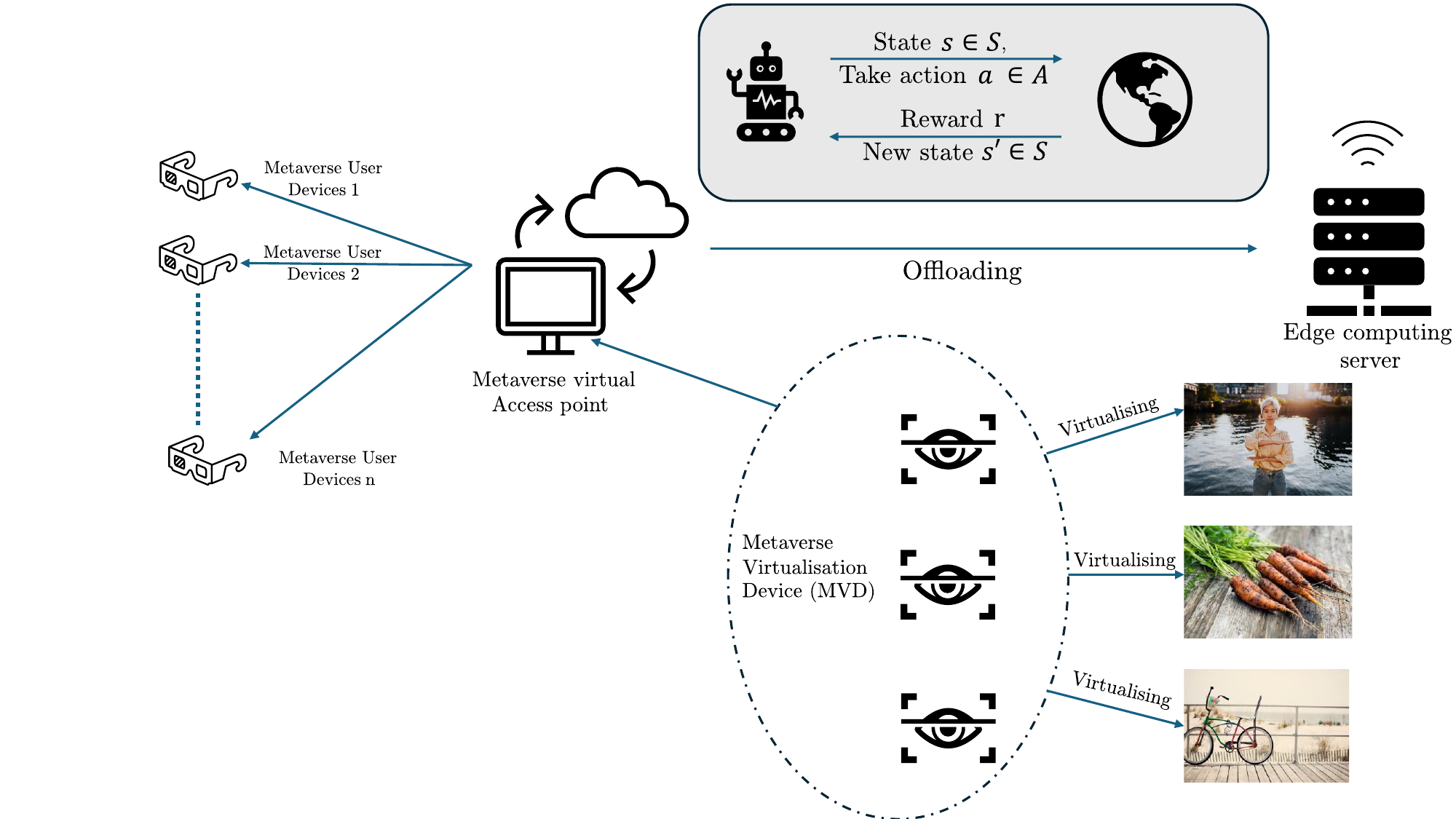}
 \caption{Model architecture}
  \label{Fig:IOT}
\end{figure*}

Our proposed model architecture, depicted in Fig.~\ref{Fig:IOT}, is built upon the architecture outlined in \cite{10122221}. We operate under the assumption that a Digital Twin (DT) manifests promptly upon one or more Metaverse user devices requesting Metaverse services, and that the DT dissipates once the Metaverse virtual access point completes serving the last Metaverse user device. Further details on the model will be elaborated in the following paragraphs.
\subsection{Model components}
\subsubsection{Collecting property from a physical system}
Metaverse Virtualisation Devices (MVD) are performed in different time slots to ensure time synchronization. We denoted $t_{n}^{\rm{Sensing}}$ and $\lambda_{n}$ as the data sensing time and sensing rate of the MVD, respectively. In this context, packets per second (PPS) is the measurement metric of $\lambda_{n}$, whereas $n_q$ bits is the number of bits in a packet. In the range of $t_{n}^{\rm{Sensing}}$, the total number of bits that a MVD collected is $n_q$$t_{n}^{\rm{Sensing}}$$\lambda_{n}$. By denoted $B_{n}$ as the quantity of received bits by MVAP from MVD $n$.  $B_{n}$ is calculated as:
\begin{equation}
B_{n}=n_qt_{n}^{\rm{Sensing}}\lambda_{n}
\end{equation}
\subsubsection{Data transmission delay}
Data transmission delay from MVD $n$ to MVAP is denoted $t^{\rm{Communication}}_{n}$, which is calculated by the equation below:
\begin{equation}
    t^{\rm{Communication}}_{n}=B_{n}/R_{n}
\end{equation}
Where $R_{n}$ is the MVD $n$ data ratio achievement and is calculated by 
\begin{equation}
\label{eq_Rn}
        R_{n} = W_{n}\log_2 \Big{(}1+ \frac{p^{\rm{Communication}}_{n}h_{n}}{\sigma^2\Gamma} \Big{)}.
\end{equation}
In this equation, we denote the bandwidth allocated to the MVD $n$, the transmit power of the MVD, the noise variance, and the channel capacity gap as $W_{n}$, $p^{\rm{Communication}}_{n}$, $\sigma^2$, and $\Gamma >1$, respectively. In the context of 6G, cell-free massive MIMO will play a significant role. Thus, we assume that the non-line of sight transition probability is very small and that the transition probability of line of sight (LoS) accounts for most of it. Therefore, we suppose only the LoS link exists between the MVD and the MVAP. Hence, $h_n$ is calculated by
\begin{equation}
    h_{n}(t) = (\frac{4\pi f}{c})^{-2} \mu_{\rm{LoS}} d_{n}^{-\beta_{\rm{LoS}}} (t) {|} {h}_{n}^{\rm{Rice}} (t) {|}^2.
\end{equation}
In which $f$ denotes the frequency of the carrier signal, $c$ denotes the speed of light, $\beta_{\text{LoS}}$ signifies the path loss exponent for the LoS component, $\mu_{\text{LoS}}$ denotes the attenuation factors related to the LoS component, $h_{n}^{\text{Rice}} (t)$ follows a Rice distribution with parameters $\nu$ and $\delta$, and $d_n$ represents the distance between MVD $n$ and the Ground Base Station (GBS).
In total, with $N$ Metaverse User Devices the amount of data bits received by MVAP is 
\begin{equation}
B^{\rm{Total}}=\sum_{n=1}^{N}B_{n}.
\end{equation}

\subsection{Process data in MVAP}
The MVAP partly processes sensing data collected from MVDs. Here, the data processing includes modeling, classification, and data cleaning. Let $B^{\rm{Offloading}} \leq B^{\rm{Total}}$ denote the amount of data bits allocated for offloading. Therefore, the number of bits that MVAP has to process on its local CPU is $B^{\rm{Local}}= B^{\rm{Total}} - B^{\rm{Offloading}}$. We denote the available CPU resource of MVAP as $f_{\rm{MVAP}}$ (cycles/second) and the processing complexity of data in cycles/bit as $\zeta$. Then, $t^{\rm{Local}}$ is determined by
\begin{equation}
t^{\rm{Local}}= \frac{\zeta}{f_{\rm{MVAP}}}B^{\rm{Local}}.
\label{local_com}
\end{equation}
\subsection{Edge Computing Server (ECS)}
We denote $\gamma_{\rm{ECS}}=\frac{g p^{\rm{Offloading}}} {I + \sigma}$ as  SINR at the ECS, where the offloading transmitting power, channel gain between ECS and MVAP, the interference power caused by the shared channel between other communication systems and MVAP, and additive thermal noise at ECS are $p^{\rm{Offloading}}$, $g$, $I$, and $\sigma$, respectively. We also denote the data rate that  MVAP achieved when choosing to offload required processing data to the ECS as $R$, and it is given by 
\begin{equation}
        R = W_{\textrm{MVAP}}\log_2 \Big{(}1+ \gamma_{\rm{ECS}} \Big{)},
\end{equation}
in which $W_{\rm{MVAP}}$ represents the bandwidth MVAP utilizes to transmit offloading bits to ECS. 

The MVAP chooses $B^{\rm{Offloading}}$ bits for offloading. Thus, the latency of communication due to the task communication $t^{\rm{Offloading}}$ is denoted by 
\begin{equation}
\label{equation_t_offloading}
t^{\rm{Offloading}}=\frac{B^{\rm{Offloading}}}{R}.
\end{equation}

By representing the time taken by the ECS to compute $B^{\rm{Offloading}}$ as $t^{\rm{ECS}}$, the overall latency induced by the offloading process is computed as follows: 
\begin{equation}
\begin{aligned}
t^{\rm{Offloading-ECS}}&=t^{\rm{Offloading}} + t^{\rm{ECS}} \\&= t^{\rm{Offloading}}+ \frac{\zeta }{f_{\rm{ECS}}}B^{\rm{Offloading}},
\end{aligned}
\end{equation}
where $f_{\rm{ECS}}$ cycles/second represents the CPU resource capacity of the ECS.
\subsection{Total Latency}
We use $t^{\rm{Delivery}}$ to denote the instances when the MVAP its services to the devices of Metaverse users devices. As the results, the total latency comprises $t_{n}^{\rm{Sensing}}$, $t^{\rm{Communication}}_{n}$, $t^{\rm{Local}}$, $t^{\rm{ECS}}$, and $t^{\rm{Delivery}}$. Because of the combination of physical system properties, it is essential to note that MVAP needs to receive all MVD data before processing it. The total latency is determined as follows:
\begin{equation}
\begin{aligned}
t^{\rm{Total}}&= t^{\rm{Sensing-Comm}} + t^{\rm{Delivery}}\\& + \max\{t^{\rm{Local}}, t^{\rm{Offloading-ECS}}\},
\end{aligned}
\end{equation}
where 
\begin{equation}
t^{\rm{Sensing-Comm}}= \max_{n\in \mathcal{N}} \{t^{\rm{Communication}}_{n} + t^{\rm{Sensing}}_{n}\}.    
\end{equation}

\subsection{Metaverse User Requirement}
This paper focuses on promptness, one of the properties mentioned in Section~\ref{sec:introduction}. Promptness is a metric that quantifies the time lag between when a feature, like the condition of roads, is observed in the real-world physical system (RWPS) and when this information becomes accessible to users in the Metaverse. We assume that the RWPS appears at time $t = 0$, and it is essential to note that to ensure the Metaveser Users' experience, the total processing time has to be smaller than the requirement time, i.e., 
\begin{equation}
\label{t_total}
    t^{\rm{Total}} \leq t^{\rm{Require}}.
\end{equation}
In addition, $t^{\rm{Require}}$ is determined by 
\begin{equation}
t^{\rm{Require}}=\min_{j \in \mathcal{J}} \{t^{\rm{Require}}_{j}\},
\end{equation}
where $j$ is $j^{th}$ the Metaverse User's devices required time and $\mathcal{J}$ is the set of ones. 
\section{Solving the requirements promptness by applying DRL }
\label{sec:DRL_offloading}
\subsection{Defining the Issue}

In this section, we present how to use the MDP defined by $<{\mathcal{S}}, {\mathcal{A}}, {\mathcal{P}r}, {\mathcal{R}}>$ to formulate the optimization of MVAP. 
\subsubsection{State Space}
The variables that vary with time and impact the objectives and decisions of the MVAP consist of $B^{\rm{Total}}$, $\gamma_{\rm{ECS}}$, $g$, $f_{\textrm{MVAP}}$, and $f_{\textrm{ECS}}$. Hence, during time slot $t$, the state space is characterized as follows:
\begin{equation}
    \mathcal{S}=\left\{ B^{\rm{Total}}, \gamma_{\rm{ECS}}, t^{\rm{Total}}, f_{\rm{MVAP}}, f_{\rm{ECS}} \right\}.
\end{equation}
In this equation, its components changed over time. First, $B^{\rm{Total}}$ alters because the channel between ECS and MVAP is dynamic and the difference in sensing rate of MVD. Next, $f_{\rm{MVAP}}$ and $f_{\rm{ECS}}$ follow the Gaussian Distribution and stochastic \cite{van2016deep}. $f_{\rm{MVAP}}$ is modeled as a normal distribution with a mean $\mu_{f_{\rm{MVAP}}}$ and a variance $\sigma_{f_{\rm{MVAP}}}^2$, represented as $f_{\rm{MVAP}} \sim \mathcal{N}(\mu_{f_{\rm{MVAP}}};\sigma_{f_{\rm{MVAP}}}^2)$. Similarly, $f_{\rm{ECS}}$ follows a normal distribution with a mean $\mu_{f_{\rm{ECS}}}$ and a variance $\sigma_{f_{\rm{ECS}}}^2$, denoted as $f_{\rm{ECS}} \sim \mathcal{N}(\mu_{f_{\rm{ECS}}};\sigma_{f_{\rm{ECS}}}^2)$. 
The randomness of the $\gamma_{ECS}$ component is determined by factors such as the communication channel between the ECS and MVAP as well as co-channel interference. A Markov chain model describes the fluctuation in SINR for $\gamma_{\rm{ECS}}$. Specifically, at time $t$, the value of $\gamma_{\rm{ECS}}$ falls within a set $\psi= \{\gamma_0,\gamma_1,...,\gamma_{\rm{S-1}} \}$ dB, where $S-1$ represents the possible values of $\gamma_{\textrm{ECS}}$. As a result, the matrix representing the probabilities of state transitions for $\gamma_{\textrm{ECS}}$ is denoted as
\begin{equation}
\Gamma_{\rm{ECS}}(t)=\left[\omega_{a_ib_i}(t)\right]_{S-1 \times S-1},
\end{equation}
where $\omega_{a_ib_i}(t) = \text{Pr}(\gamma(t)=b_i\Big{|} \gamma(t)=a_i)$ with $a_i,b_i\in \psi$. Here, $\omega_{a_ib_i}(t)$ signifies the likelihood of $\gamma_{\textrm{ECS}}$ transitioning from $a_i$ to $b_i$ between time $t$ and $t+1$.
\subsubsection{Action Space}
The MVAP determines the amount of data to be offloaded, denoted as $B^{\rm{Offloading}}$. The set of possible actions $\mathcal{A}$ is defined as follows:
\begin{equation}
\mathcal{A}=\left\{B^{\rm{Offloading}} \Big{|} B^{\rm{Offloading}} \in \mathbf{W}\right\},
\end{equation}
where $\mathbf{W}$ presents $\left[0;\frac{B^{\rm{Total}}}{F};\cdots;\frac{(F-1)B^{\rm{Total}}}{F};B^{\rm{Total}} \right]$,  $B^{\rm{Offloading}}$ ranges from 0, indicating that the MVAP will handle the entire task internally, to $B^{\rm{Total}}$, indicating that the MVAP will offload the entire data size to the ECS for processing. Here, $F$ represents the task-splitting factor. It is important to note that this division of tasks may result in a non-integer task size. In such cases, the MVAP, acting as the agent, will round the value up to the nearest integer.

\subsubsection{Reward}
The reward structure is designed to ensure compliance with the user latency requirements outlined in (\ref{t_total}), which specifies that the total latency should not exceed the defined threshold for Metaverse user devices. Specifically, the MVAP is rewarded positively if $t^{\rm{Total}}$ (total latency) is less than or equal to $t^{\rm{Require}}$ (latency requirement) and negatively if $t^{\rm{Total}}$ exceeds $t^{\rm{Require}}$. The positive reward is denoted as $Positive-value$, and the negative reward is denoted as $Negative-value$. The instant reward received by the MVAP, $r$, is defined as follows:
\begin{align} {r=}
    \begin{cases}
        Positive-value \text{ if }   t^{\rm{Total}} \leq t^{\rm{Require}} \\
        Negative-value\text{ if }   t^{\rm{Total}} > t^{\rm{Require}}.
    \end{cases}
\label{eq:labelone}
\end{align}
Positive and negative values can generally be distinguished and determined through empirical observations. The primary goal of the MVAP is to optimize the average long-term reward by identifying an optimal policy, denoted as $\pi^*: {\mathcal{S}} \rightarrow {\mathcal{A}}$. This policy enables the MVAP to select the most suitable action $a \in {\mathcal{A}}$ for any given state $s \in {\mathcal{S}}$.
\subsection{Algorithms}
The Deep Q-Network (DQN) and Double Deep Q-learning Network (DDQN) are two reinforcement learning algorithms used to learn and optimize the behavior of an agent within an environment modeled by a Markov Decision Process (MDP). These algorithms are detailed in Algorithms~\ref{DQN} and~\ref{DDQN}, with their parameters illustrated in Table~\ref{Para_table}.

DQN employs a neural network (Q-network) to estimate Q-values for state-action pairs in an MDP. It uses techniques like experience replay and a target network to stabilize and improve learning. The target network in DQN is periodically updated to minimize the estimation errors of Q-values.

DDQN extends DQN by utilizing two separate Q-networks: one for action selection (primary Q-network) and another for value estimation (target Q-network). DDQN addresses the overestimation bias often seen in DQN by decoupling action selection from value estimation. It updates the primary Q-network more frequently while periodically updating the target Q-network to reflect improvements, leading to more stable and accurate learning.

In summary, both DQN and DDQN aim to learn optimal policies by approximating Q-values, with DDQN specifically designed to mitigate overestimation bias and enhance learning stability.

For more details, we present DQN and DDQN in algorithms~\ref{DQN} and~\ref{DDQN}. The computational complexity of training the DDQN algorithm can be expressed as $\mathcal{O}(TN_b (|La_0||La_1| + |La_1||La_2| + |La_2||La_3| + |La_3||La_4|))$. Here, $N_b$ represents the batch size during training, and $T$ indicates the total number of training iterations. In our study, the dimension of $La_0$ represents the number of features describing the state, denoted as $|La_0|=5$. The hidden layers $La_1$, $La_2$, and $La_3$ are equal to 256 neurons. The outcome layer $La_4$ contains $F$ neurons, each corresponding to an action available to the agent; in paper context, $F=1000$.
\begin{algorithm}
\caption{Deep Q-Network (DQN)}
\label{DQN}
\begin{algorithmic}[1]
    \State Establish: $\mathcal{D}$ with $N$, $\theta$, $\theta' = \theta$, $\gamma$, $\epsilon$, $ Q^{\textrm{pri}}_{\textrm{net}}$, and $Q^{\textrm{tar}}_{\textrm{net}}$
    \For{episode $= 1$ to len(episode)}
        \State Establish $s_1$
        \For{timestep $t = 1$ to $T$}
            \State Based on $\epsilon$ choose $a_t$
            \State Otherwise, choose $a_t = \text{argmax}_a Q(s_t, a; \theta)$
            \State Do $a_t$ and consider $r_t$ and $s_{t+1}$
            \State By utilising $\mathcal{D}$ to store $(s_t, a_t, r_t, s_{t+1})$
            
            \State From $\mathcal{D}$, sample random minibatch $(s_i, a_i, r_i, s_{i+1})$ 
            \State By using $Q^{\textrm{tar}}_{\textrm{net}}$, calculate $Q^{\textrm{tar}}_{\textrm{val}}$:
            \[
            y_i = 
            \begin{cases}
            r_i & \text{if * } \\
            r_i + \textbf{Equation}(~\ref{dqn_Q})& \text{otherwise}
            \end{cases}
            \]
            \State Update $Q^{\textrm{val}}_{\textrm{net}}$ parameters $\theta$ using gradient descent with loss:
            \[
            \mathcal{L}(\theta) = \frac{1}{B} \sum_{i=1}^{B} \left( Q(s_i, a_i; \theta) - y_i \right)^2
            \]
            \State Every $C$ timesteps: $\theta' =\tau \theta + (1-\tau) \theta'$
            \State  $\epsilon = \epsilon \times \epsilon_{\text{decay}}$
        \EndFor
    \EndFor
\end{algorithmic}
\begin{equation}
    \label{dqn_Q}
    \gamma \max_{a'} Q(s_{i+1}, a'; \theta')
\end{equation}
\end{algorithm}
\begin{algorithm}
\caption{Double Deep Q-Learning (DDQN)}
\label{DDQN}
\begin{algorithmic}[1]
    \State Establish: $\mathcal{D}$ based on $N$, $\theta$, $\theta'$, $C$,  $\gamma$, $\epsilon$, $\epsilon_{\text{decay}}$, $ Q^{\textrm{pri}}_{\textrm{net}}$, and $Q^{\textrm{tar}}_{\textrm{net}}$
    \For{episode $= 1$ to len(episode)}
        \State Establish $s_1$
        \For{$t = 1$ to $T$}
            \State Based on $\epsilon$ choose $a_t$
            \State Otherwise, choose $a_t = \text{argmax}_a Q(s_t, a; \theta)$
            \State Do $a_t$ and consider $r_t$ and $s_{t+1}$
            \State By utilising $\mathcal{D}$ to store $(s_t, a_t, r_t, s_{t+1})$
            \State From $\mathcal{D}$, sample random minibatch $(s_i, a_i, r_i, s_{i+1})$ 
            \State By using the $Q^{\textrm{tar}}_{\textrm{net}}$, calculate $Q^{\textrm{tar}}_{\textrm{val}}$:
            \[
            y_i = 
            \begin{cases}
            r_i & \text{if *} \\
            r_i + \textbf{equation} ( \ref{Q_DDQN} ) & \text{otherwise}
            \end{cases}
            \]
            
            \State Update $Q^{\textrm{pri}}_{\textrm{net}}$ parameters $\theta$ using gradient descent with loss:
            \[
            \mathcal{L}(\theta) = \frac{1}{B} \sum_{i=1}^{B} \left( Q(s_i, a_i; \theta) - y_i \right)^2
            \]
            \State Every $C$ timesteps $\theta' = \tau \theta + (1-\tau) \theta'$
            \State  $\epsilon = \epsilon \times \epsilon_{\text{decay}}$
        \EndFor
    \EndFor
\end{algorithmic}
\begin{equation}
    \label{Q_DDQN}
    \gamma Q(s_{i+1}, \text{argmax}_a Q(s_{i+1}, a; \theta); \theta')
\end{equation}
\end{algorithm}
\begin{table}[t]
\centering
\caption{Parameter explanation of algorithm Q-LEARNING to ~\ref{DDQN}}
\label{Para_table}
\begin{tabular}{|l|l|}
\hline
Parameter & Definition \\ \hline
$\mathcal{D}$         & replay memory           \\ \hline
 $N$         & capacity           \\ \hline
 $\theta$         & weights of primary Q-network          \\ \hline
 $\theta'$         & weights of target Q-network           \\ \hline
  $C$        & frequency of updating target network       \\ \hline
 $\gamma$         & discount factor         \\ \hline
  $\epsilon$        & exploration rate         \\ \hline
  $\epsilon_{\textrm{decay}}$        & decay rate         \\ \hline
  $B$        &  minibatch size         \\ \hline
  $\tau$        &  update rate          \\ \hline
  $*$        &  episode terminates at step $i + 1$         \\ \hline
  $s_i$        & state of agent at $i^{th}$ step      \\ \hline
  $r_i$        & reward of agent at $i^{th}$ step      \\ \hline
  $a_i$        & action of agent at $i^{th}$ step      \\ \hline
  $s_{i+1}$        & state of agent at ${(i+1)}^{th}$ step      \\ \hline
    $s_t$        & state of agent at $i^{th}$ episode      \\ \hline
  $r_t$        & reward of agent at $i^{th}$ episode      \\ \hline
  $a_t$        & action of agent at $i^{th}$ episode      \\ \hline
  $s_{t+1}$        & state of agent at ${(i+1)}^{th}$ episode      \\ \hline
   $Q^{\textrm{pri}}_{\textrm{net}}$        & prior Q-network    \\ \hline
    $Q^{\textrm{tar}}_{\textrm{net}}$        &  target Q-network       \\ \hline
       $Q^{\textrm{pri}}_{\textrm{val}}$        & prior Q-value    \\ \hline
    $Q^{\textrm{tar}}_{\textrm{val}}$        &  target Q-value       \\ \hline
       $a$        & current action    \\ \hline
       $a'$        & next action    \\ \hline
    $r$        & reward       \\ \hline
       $s$        & current state    \\ \hline
    $s'$        &  next state       \\ \hline
\end{tabular}
\end{table}

\section{EXPERIMENT RESULTS}
This section presents the numerical findings used to assess the performance of the proposed DRL algorithm. To establish baselines, we compare the performance of the DDQN algorithm against the DQN and Q-learning (QL) algorithms. For Q-learning, we discretized the data to make the algorithm implementable, creating discrete data for the state space. Additionally, we include the ``RM'' baseline, which indicates the average reward obtained when the MVAP selects actions randomly without following a specific policy. The value of packet length $q$ is randomly taken from a uniform distribution $[960 \times 540, 1280 \times 720, 1920 \times 1080, 2560 \times 1440]$ bits/packet. For $\gamma_{\rm ECS}$, we set $\psi= \{-5,-3, 0,3, 5 \}$ dB and its state transition probability matrix based on Markov chain is defined as
\begin{equation*}
 \Gamma_{\rm ECS} =  \begin{bmatrix}
0.600 & 0.250 & 0.100 & 0.040 & 0.010\\
0.250 & 0.300 & 0.250 & 0.100 & 0.040\\
0.100 & 0.250 & 0.320 & 0.250 & 0.100\\
0.040 & 0.100 & 0.250 & 0.300 & 0.250\\
0.010 & 0.040 & 0.100 & 0.250 & 0.600
    \end{bmatrix}.
\end{equation*}

We set $Negative-value = -1 $  and $Positive-value = 20$ to encourage the agent of the MVAP to choose the appropriate course of action. It is essential to note that if we set the $Negative-value = -1 $ too close to the $Positive-value$, the agent will not learn to obtain the better optimal reward. Table~\ref{tab:my_label} contains a list of additional simulation settings for the algorithms and network model.
\begin{table}[h]
\scriptsize
\centering
\caption{PARAMETERS USED TO SIMULATE PROPOSED MODEL}
\begin{tabular}{|c|c|c|c|}
\hline
Parameter & Parameter value & Parameter & Parameter value \\
\hline
$f_{\rm{MVAP}}$ & $\sim 10.5$ GHz & $N$ & $3$ \\
$f_{\rm{ECS}}$ & $\sim 20.5$ GHz & $I$ & $3$ \\
$t^{\rm{Require}}_{j}$ & $\mathcal{U} [1.5, 2.3]$ second & $d_{n}$ & $160-210$ (m) \\
$p^{\rm{Communication}}_{n}$ & $0.52$ Watts & $\zeta$ & $650$ cycles/bit \\
$\sigma^2$ & $10^{-11}$ Watts & $t^{\rm{Sensing}}_n$ & $0.5$ second \\
$\beta_0$ & $10^{-6}$ \cite{8062796} & $\lambda_{n}$ & $5$ packets/second \\
$\gamma$ & $2.2$ \cite{8062796} & Learning rate & $0.001$ \\
$\Gamma$ & $1.2$ \cite{8062796} & Batch size & $10$ \\
$\epsilon$-greedy & $1 \to 0.001$ & Discount rate & $0.9985$ \\
$\mu_{f_{\rm{MVAP}}}$ & $10.5$ GHz & $\mu_{f_{\rm{ECS}}}$ & $20.5$ GHz \\
$\sigma_{f_{\rm{MVAP}}}^2$ & $0.1 \mu_{f_{\rm MVAP}}$ & $\sigma_{f_{\rm{ECS}}}^2$ & $0.1 \mu_{f_{\rm{ECS}}}$ \\
\hline
\end{tabular}
\label{tab:my_label}
\end{table}

\begin{figure}[t]
    \includegraphics[width=\linewidth]{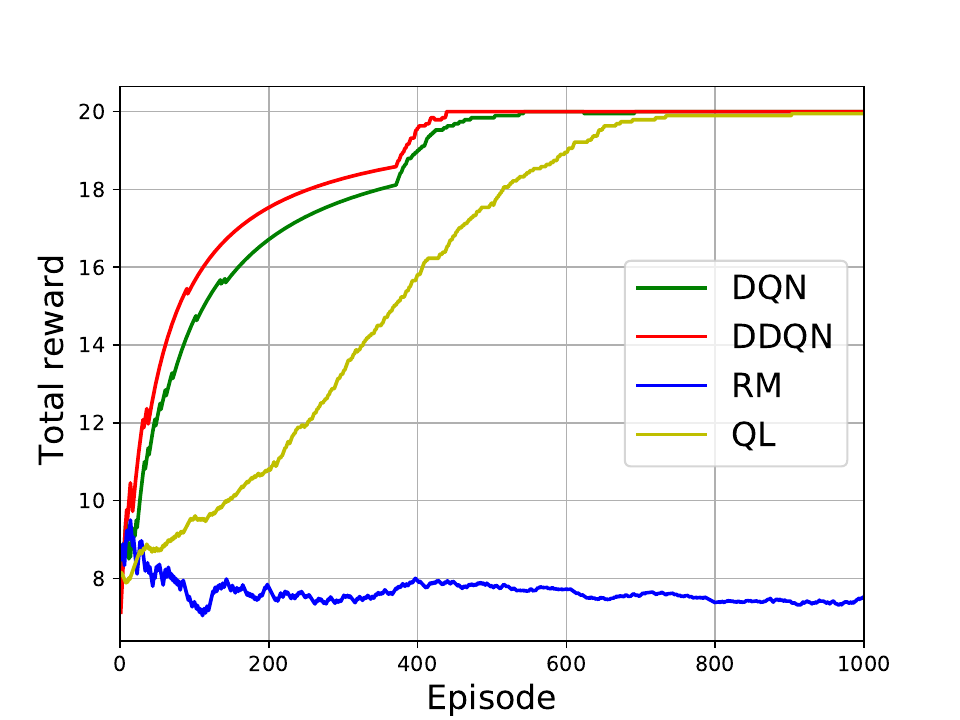}
     \caption{Algorithm convergence, 3 MVDs}
    \label{fig:Convergence}
\end{figure}
 Fig.~\ref{fig:Convergence} illustrates that the DDQN, DQN, and QL algorithms converge to maximum reward values. The rewards of these algorithms are remarkably more elevated than the reward received by the RM. Additionally, DDQN converges faster than DQN at around 450 and 650 episodes due to decoupling action selection and evaluation tasks. Q-learning converges after around 950 episodes. It is shown that the Q-table of ones is updated extremely slowly to find optimal action in a dynamic environment. RM shows the worst average reward, which is under 10. Additionally, Fig.~\ref{fig:Convergence} shows the $t^{\textrm{Require}}$ randomly from 1.4 to 2.3 seconds, $t^{\textrm{Sensing}}=0.5$ seconds with the number of Metaverse user devices is 3, the QL still can coverage at the reward of nearly 20.
\begin{figure}[!h]
    \centering
    \includegraphics[width=\linewidth]{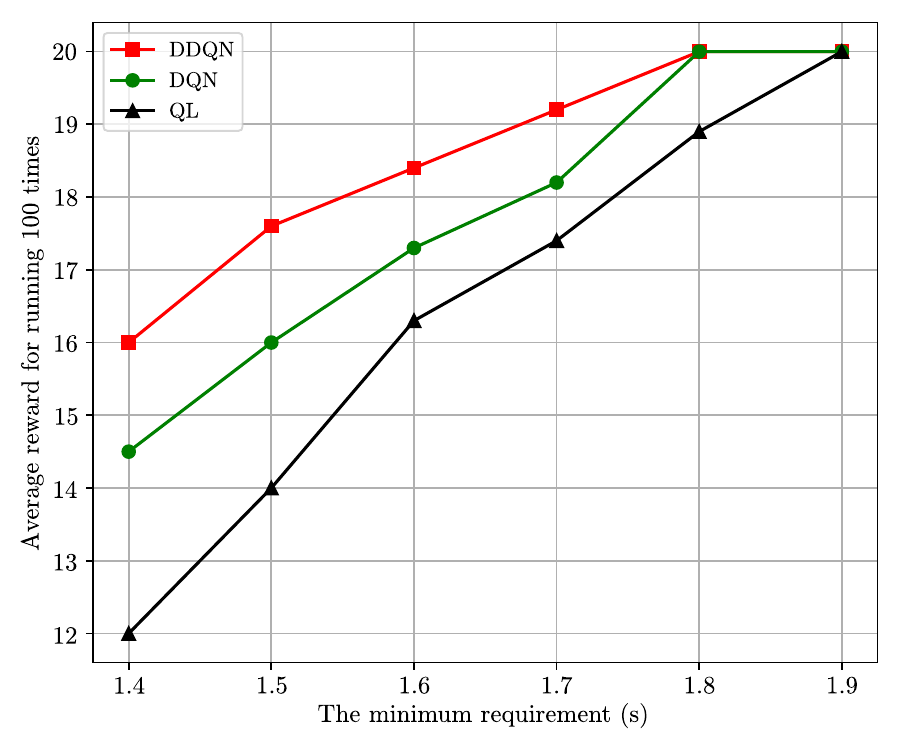}
    \caption{The requirement of latency of MU}
    \label{fig:reward_complex}
\end{figure}

It is essential to talk about how Metaverse awards are affected by the minimum latency requirement $t^{\rm{Require}}$. Fig.\ref{fig:reward_complex} illustrates how rewards for all algorithms rise with $t^{\rm{Require}}$. This is because (\ref{eq:labelone}) has a more easily satisfied latency requirement, which gives the MVAP higher positive ratings. Nevertheless, the rewards will not increase beyond a certain latency threshold. If $t^{\rm{Require}}$ exceeds 1.8 seconds, the rewards for both the DQN and DDQN algorithms remain unchanged, as the latency condition is consistently met.

\section{Conclusion}
We investigated a Metaverse system with edge computing support, where an MVAP optimizes the data bits offloaded to the Edge Server (ES) to ensure timely Metaverse services that meet consumers' latency requirements. Due to the system's random and dynamic nature, we defined the MVAP's offloading problem as a stochastic problem. To address this issue, we implemented DRL methods, such as DDQN and DQN. The simulation results demonstrate the effectiveness of the proposed algorithms. Future research could explore scenarios involving multiple MVAPs, where advanced learning algorithms like multi-agent DRL might prove beneficial.





\bibliographystyle{IEEEtran}
\bibliography{IEEE}
\end{document}